\documentclass[aps,prd,amsmath,floats,floatfix,twocolumn,
superscriptaddress,nofootinbib,showpacs,longbibliography,reprint]{revtex4-2}

\usepackage[T1]{fontenc}
\usepackage[utf8]{inputenc}
\usepackage{lmodern}

\usepackage{verbatim}

\usepackage[dvipsnames, usenames]{xcolor}
\definecolor{linkcolor}{rgb}{0.0,0.3,0.5}
\usepackage[hypertexnames=false, unicode, colorlinks=true, linkcolor=linkcolor,
citecolor=linkcolor,filecolor=linkcolor,urlcolor=linkcolor,]{hyperref}
\usepackage{etoolbox}
\usepackage[all]{hypcap}
\usepackage{graphicx}
\usepackage{xspace}
\usepackage{amssymb}
\usepackage[normalem]{ulem} %for \sout
\usepackage{bm} % boldmath

\usepackage{microtype}

\usepackage[english]{babel}
\usepackage{blindtext}

\usepackage[caption=false]{subfig}

\usepackage{orcidlink}

\graphicspath{%
  {figs/}%
}

\DeclareMathAlphabet{\mathpzc}{OT1}{pzc}{m}{it}

\newcommand{\enclosepar}[1]{\left( #1 \right)}

\newcommand{\enclosebar}[1]{\left| #1 \right|}

\begin{document}

\title{Eccentricity Reduction for Quasicircular Binary Evolutions}

\newcommand{\Cornell}{\affiliation{Cornell Center for Astrophysics and Planetary
    Science, Cornell University, Ithaca, New York 14853, USA}}
    \newcommand\CornellPhys{\affiliation{Department of Physics, Cornell
    University, Ithaca, New York 14853, USA}}
    \newcommand\Caltech{\affiliation{TAPIR 350-17, California Institute of
    Technology, 1200 E California Boulevard, Pasadena, CA 91125, USA}}

\author{Sarah Habib\,\orcidlink{0000-0002-4725-4978}}
\email{shabib@caltech.edu}
\Caltech

\author{Mark A. Scheel\,\orcidlink{0000-0001-6656-9134}}
\Caltech

\author{Saul A. Teukolsky\,\orcidlink{0000-0001-9765-4526}}
\Caltech
\Cornell

% Because hyperref only gets the *last* author, we need to be explicit.
\hypersetup{pdfauthor={Habib et al.}}

\date{\today}

%==========================================================================
\begin{abstract}
Simulation of quasicircular compact binaries is a major goal in numerical
relativity, as they are expected to constitute most gravitational wave
observations. However, given that orbital eccentricity is not well-defined in
general relativity, providing initial data for such binaries is a challenge for
numerical simulations. Most numerical relativity codes obtain initial conditions
for low-eccentricity binary simulations by iterating over a sequence of short
simulations --- measuring eccentricity mid-evolution and correcting the initial
data parameters accordingly. Eccentricity measurement depends on a numerically
challenging nonlinear fit to an estimator model, and the resulting eccentricity
estimate is extremely sensitive to small changes in how the fit is performed. We
have developed an improved algorithm that produces more consistent measurements
of eccentricity relative to the time window chosen for fitting. The primary
innovations are the use of the nonlinear optimization algorithm, variable
projection, in place of more conventional routines, an initial fit parameter
guess taken from the trajectory frequency spectrum, and additional frequency
processing of the trajectory data prior to fitting.
\end{abstract}

\maketitle

%==========================================================================
\section{Introduction}
\label{sec:introduction}

Quasicircular compact binaries are expected to make up the majority of
binaries detected by gravitational wave observatories. This is because
gravitational radiation causes binaries
to circularize during inspiral,
resulting in low eccentricity at merger
\cite{Peters:1963, Peters:1964}.  Indeed, most studies of gravitational-wave
(GW) detection events show that the waveforms are consistent with
quasicircular
orbits~\cite{Romero-Shaw:2020thy,Wu:2020zwr,Gayathri:2020coq,Romero-Shaw:2021ual,OShea:2021faf,Gamba:2021gap,Iglesias:2022xfc,Ramos-Buades:2023yhy}. However,
there is evidence that some events may have nonzero
eccentricity~\cite{Gupte:2024jfe}. This suggests that some binaries
might have formed relatively recently by dynamical processes,
for example, in dense environments such as
globular
clusters~\cite{OLeary:2005vqo,Samsing:2017xmd,Fragione:2018vty,Zevin:2021rtf}.

Our best understanding of GW events from binary mergers relies on
numerical relativity (NR). This can be either from
direct simulations, or from surrogate
models~\cite{Varma:2019csw,Boschini:2023ryi,Yoo:2023spi,Islam:2022laz,Yoo:2022erv,Islam:2021mha}
or analytical waveform
models~\cite{Cotesta:2018fcv,Mehta:2017jpq,London:2017bcn}
calibrated to NR.  These NR simulations are formulated in terms of
initial-value problems that involve two steps. The first step is the
construction of \emph{initial data} \cite{Cook:2000vr} that satisfy the Einstein
constraint equations on some surface of constant coordinate time. In
this step properties such as the masses and spins of the objects and their
initial positions and velocities are freely
specifiable.  The second step is the \emph{evolution} of the initial
data through time, which yields the spacetime metric as a function of
time, including the emitted gravitational radiation.

To understand quasicircular binary inspirals, it is therefore
important to construct NR simulations that have nearly zero orbital
eccentricity. In Newtonian physics this would be straightforward:
Kepler's laws allow specifying initial positions and velocities of the
objects that correspond to zero eccentricity. However, this is not
true in general relativity (GR).  First of all, in GR there are no
truly circular orbits because of radiation reaction. Thus, the goal is to
achieve a quasicircular orbit in which the binary orbit decays at a
monotonic (as opposed to oscillatory) rate, in the absence of spin.
For spinning objects, the goal is still to reduce oscillations in the
orbit, but this is more complicated because spin-spin interactions produce
oscillations that must be distinguished from those caused by
eccentricity~\cite{Buonanno:2010}.  In the post-Newtonian (PN)
approximation it is possible to compute expressions for particle
positions and velocities (or equivalently, initial orbital separation,
orbital frequency, and radial velocity) that give quasicircular
orbits~\cite{Blanchet:2006zz}, and indeed by using high enough order
PN these expressions can be
used~\cite{Healy:2017zqj,Ciarfella:2024clj} to produce NR simulations with
eccentricity of order $10^{-3}$. However, for accurate waveforms we
want to achieve eccentricities smaller than this.

One potential drawback of using PN expressions directly is that the
gauge (coordinate) choices used in PN typically differ from those used in
NR~\cite{Pfeiffer:2002xz,Varma:2018sqd}, so that PN and NR orbits can
disagree to an extent that is difficult to predict. Yet another
problem stems from initial transients that appear in NR simulations.
These transients occur because NR initial data, even in the
infinite-resolution limit, does not contain the same gravitational
radiation as would a snapshot of a binary that has been inspiralling
since the infinite past.  Instead, at the beginning of the evolution
the solution quickly relaxes to quasiequilibrium, slightly changing
the initial parameters and the orbit and emitting
high-frequency gravitational waves in the process; these waves are
known as ``junk radiation.''  There have been attempts to reduce the
amount of junk radiation in NR
simulations~\cite{Alvi:1999cw,Yunes:2006iw,JohnsonMcDaniel:2009dq,Kelly:2009js,Reifenberger:2012yg,Tichy:2016vmv,Lovelace:2008hd,Varma:2018sqd},
but typically one simply discards the first few orbits of the
simulation until the junk radiation has decayed
away~\cite{Boyle:2019}.

\begin{figure}[t]
    \centering
    \includegraphics[width=\columnwidth]{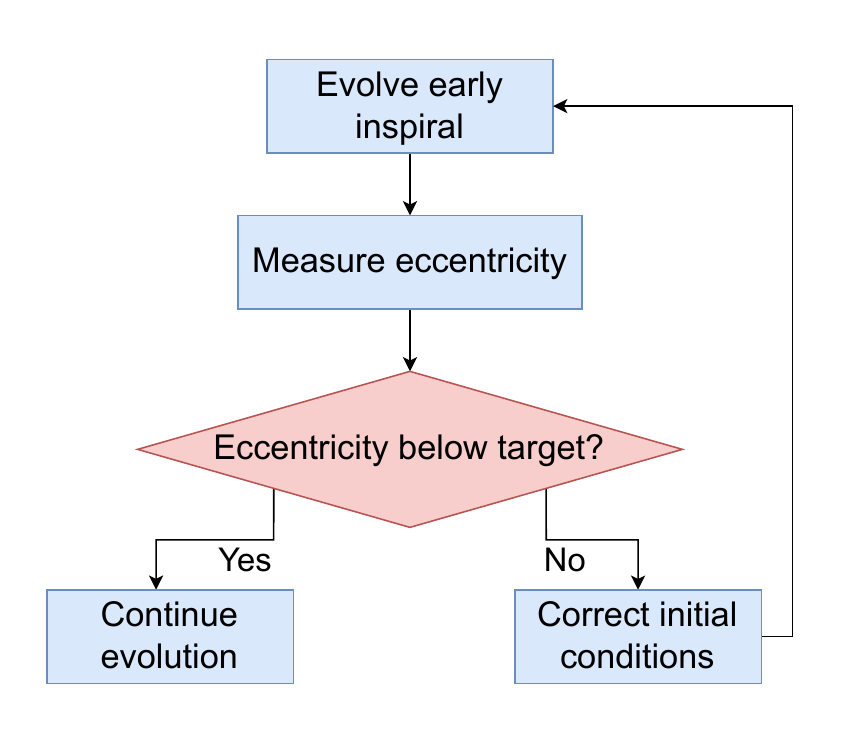}
    \caption{A simplified schematic of how quasicircular binary evolutions are
    achieved in NR codes. After the inspiral has run for
    1 -- 2 orbital periods after gravitational wave junk radiation, the
    trajectory is fit to the model in Eq.~\eqref{eq:omegadot_impl} and an
    estimate of the eccentricity is calculated as a derived quantity from the
    fit parameters. If the eccentricity is below a set target value, the
    evolution is continued. Otherwise, the initial conditions are updated using
    the formulation in Eq.~\eqref{eq:update} and a new evolution is started.
    This cycle of restarting evolution with updated initial conditions typically
    needs to be repeated multiple times in order to achieve an
    eccentricity of order $10^{-4}$.}
    \label{fig:flowchart}
\end{figure}

To handle the above difficulties, we have adopted an iterative method
for producing quasicircular NR initial
data~\cite{Pfeiffer:2007,Buonanno:2010,Purrer:2012,Ramos-Buades:2018}: An initial guess is chosen
for NR initial data parameters, the binary is evolved for a few orbits
using NR and the eccentricity is measured from that evolution, and
then the initial guess is updated so as to give smaller eccentricity
for the next iteration. See Fig.~\ref{fig:flowchart} for an
illustration of the process.

In this paper we discuss improvements to the iterative
eccentricity-reduction method of Ref.~\cite{Buonanno:2010} and its
implementation in the NR code \texttt{SpEC}~\cite{SpECwebsite}.
Although we limit the discussion and particular examples to black-hole
binaries and to \texttt{SpEC}, the methods here can also be
used for binaries containing neutron stars and in other NR codes.

The improvements discussed here deal with the eccentricity measurement
stage of Fig.~\ref{fig:flowchart}.  This stage involves extracting
coordinate trajectories of BHs from an NR simulation, computing the
orbital angular velocity $\Omega(t)$ from the trajectories, fitting
its time derivative $\dot\Omega(t)$ to a PN-inspired formula
(Eq.~\eqref{eq:omegadot_impl} below) that involves the eccentricity,
and reading off the eccentricity from the fit.  Before our
improvements, the implementation of this procedure was not very
robust. For example, small changes in the time interval over which
$\dot\Omega(t)$ is fit or small changes in the initial guesses for the
fit parameters often led to large changes in the measured eccentricity.
Occasionally, the fitting procedure completely failed to converge.
We show below that our changes greatly improve the robustness of the
algorithm.

This paper is organized as follows. In Sec.~\ref{sec:methods}, we
first describe how eccentricity reduction is currently performed in the
\texttt{SpEC} code, and then present several new techniques to address the
shortcomings of the current method. In Sec.~\ref{sec:results}, we
apply these techniques to the SXS public waveform
catalog~\cite{Boyle:2019} and compare the consistency of measured
eccentricities with that of the current eccentricity reduction method.
Finally, we summarize our findings in Sec.~\ref{sec:conclusion}. We
adopt the unit convention $G=c=1$.

%==========================================================================
\section{Methods}
\label{sec:methods}

\subsection{Current eccentricity reduction method}
\label{sec:eccred}

Here we outline the method that is currently used for eccentricity reduction in
\texttt{SpEC}. This method is described in more detail in
Ref.~\cite{Buonanno:2010}.

To start a binary evolution at some initial time, one must specify the positions
and the velocities of the objects at that time.  Without loss of generality, we
assume the two objects are initially on the positive and negative $x$ axis
with initial positions $\vec{x}_1(0)$ and $\vec{x}_2(0)$, and their
initial velocities $\dot{\vec{x}}_1(0)$ and $\dot{\vec{x}}_2(0)$ are in the $xy$
plane.  We then specify the initial coordinate
distance $r_0 = \vec{x}_1(0) - \vec{x}_2(0)$
between the objects, the time derivative of this distance $\dot{r}_0 =
\dot{\vec{x}}_1(0) - \dot{\vec{x}}_2(0)$, and the initial orbital frequency
$\Omega_0$ of the binary. Given the separation $\vec{r}(t) = \vec{x}_1(t) -
\vec{x}_2(t)$ of the binary, orbital frequency can be defined as
\begin{align}
  \Omega(t) = \frac{\enclosebar{\vec{r} \times \dot{\vec{r}}}}{r^2}
\end{align}

For a Newtonian binary, a circular orbit would be achieved by setting
$\dot{r}_0=0$ and setting $r_0$ and $\Omega_0$ according to Kepler's law. For a
relativistic binary, there are no true circular orbits because of radiation
reaction. One can write post-Newtonian (PN) expressions for a choice
of $\dot{r}_0$, $r_0$, and $\Omega_0$ that achieves a quasicircular orbit
\cite{Blanchet:2006zz,Healy:2017zqj,Ciarfella:2024clj}, but these
expressions do not account for possible gauge differences between
PN and NR.

Therefore, as described above,
 we use the iterative procedure summarized in
Figure~\ref{fig:flowchart}: We guess values of $\dot{r}_0$,
$r_0$, and $\Omega_0$ (typically using low-order PN), evolve
for a few binary orbits, and measure the eccentricity.
We then use this measurement to update the values of $\dot{r}_0$, $r_0$, and
$\Omega_0$ for the next iteration.  The procedure stops when the eccentricity is
below some tolerance, at which point we evolve to the desired final time.

For each step in the iterative procedure, initial data is generated from the
given $r_0$, $\dot{r}_0$, and $\Omega_0$ values using an initial data solver
such as \texttt{Spells} \cite{Pfeiffer:2003}. The initial data is numerically
evolved through early inspiral, and the eccentricity is then measured using the
extracted time series of orbital frequency ${\Omega}(t)$ and its first time
derivative $\dot{\Omega}(t)$. These quantities are computed from the coordinate
trajectories of the centers of the apparent horizons. An alternative would be to
compute the eccentricity from the gravitational waveform. However, using the
coordinate trajectories has the advantage of requiring fewer orbits (and less
simulation run time) than a method based on waveform data. Using waveform data
would require running the simulation long enough for the signal to propagate to
a region far from the source where the waveform can be measured. We expect that
the difference between eccentricity measured from the trajectory and measured
from the waveform to be unimportant for the purposes of eccentricity reduction.

For eccentricity estimation and corrections to the initial data parameters, we
use the formulation described
in Ref.~\cite{Buonanno:2010}; for clarity, we
will outline it here.

Updates to the initial data parameters require three quantities that are not
directly output by the simulation: initial eccentricity $e$, frequency of
eccentricity-induced oscillations $\omega$, and initial mean anomaly $\phi_0$.
These three quantities
are estimated by fitting a phenomenological model to the time
derivative of the orbital frequency.

In the derivation of updating formulae, we follow the $\dot{\Omega}(t)$ model of
Ref.~\cite{Buonanno:2010}. Newtonian mechanics gives for the
relation between $\dot{\Omega}$ and $e$
\begin{align}
    \label{eq:Newtonian_omega}
    \Omega(t) & = \bar{\Omega} + 2e\bar{\Omega}\sin\enclosepar{\omega t +
                \phi_0}, \\
    \dot{\Omega}(t) & = 2e\bar{\Omega}\omega\cos\enclosepar{\omega t + \phi_0},
\end{align}
where $\Omega(t)$ is the time-dependent orbital frequency
and $\bar{\Omega}$ is the mean value of orbital frequency (i.e., without
eccentricity or spin-induced oscillations). Including additional terms for
radiation reaction and spin-spin interaction, the eccentricity estimator model
becomes
%\begin{align}
%    \label{eq:omegadot_analytic}
%    \dot{\Omega}(t) &= S_\Omega(t) + 2 e \bar{\Omega}\omega\cos(\omega t +
%                      \phi_0) \nonumber \\
%  &- \frac{\bar{\Omega}}{M^2 \bar{r}} F \sin(2\bar{\Omega} t
%  + \gamma),
%\end{align}
\begin{equation}
    \label{eq:omegadot_analytic}
    \dot{\Omega}(t) = S_\Omega(t) + 2 e \bar{\Omega}\omega\cos(\omega t +
                      \phi_0)
  - \frac{\bar{\Omega}}{M^2 \bar{r}} F \sin(2\bar{\Omega} t
  + \gamma),
\end{equation}
where $S_\Omega(t)$ characterizes the non-oscillatory increase in mean value of
$\dot{\Omega}(t)$ resulting from radiative energy loss during inspiral,
$M=m_1+m_2$ is the total mass of the  binary, and $\bar{r}$ is the mean value of
separation. The last term in Eq.~\eqref{eq:omegadot_analytic} accounts for
eccentricity-independent oscillations in $\dot\Omega(t)$ caused by spin-spin
interactions. The quantity $F$ is defined by
\begin{align}
    \label{eq:spin_coefficient}
    F = (\vec{S}_0 \cdot \hat{n}_0)^2 + (\vec{S}_0 \cdot
  \hat{\lambda}_0)^2,
\end{align}
 where $\vec{S}_0 = \enclosepar{1 + \frac{m_2}{m_1}}\vec{S}_1 + \enclosepar{1 +
 \frac{m_1}{m_2}}\vec{S}_2$ is the initial total spin vector, and
 \begin{align}
    \label{eq:triad}
    \hat{n}(t) = \frac{\vec{r}}{r}, \ \hat{L}_N(t) = \frac{\vec{r} \times
   \dot{\vec{r}}}{\enclosebar{\vec{r} \times \dot{\vec{r}}}}, \ \hat{\lambda}(t) =
   \hat{L}_N \times  \hat{n}, \\
   \label{eq:triad_0}
   \hat{n}_0 = \hat{n}(0), \ \hat{L}_{N0} = \hat{L}_N(0), \ \hat{\lambda}_0 = \hat{\lambda}(0),
\end{align}
so $\hat{L}_N$ is orthogonal to the instantaneous orbital plane. The quantity
$\gamma$ is defined such that
 \begin{align}
    \label{eq:gamma}
    \sin\gamma & = \cos\alpha\sin\alpha, \\
    \label{eq:alphadef}
    \cos\alpha & \equiv \frac{\vec{S}_0 - \enclosepar{\vec{S}_0 \cdot
                 \hat{L}_{N0}}\hat{L}_{N0}}{\enclosebar{\vec{S}_0 - \enclosepar{\vec{S}_0 \cdot
                 \hat{L}_{N0}}\hat{L}_{N0}}} \cdot \hat{n}_0,
 \end{align}
where Eq.~\eqref{eq:alphadef} defines $\alpha$. Given that the final term in
Eq.~\eqref{eq:omegadot_analytic} is included only to capture spin-induced
effects, the second term captures the full contribution to $\dot{\Omega}(t)$
from eccentricity.
The goal of eccentricity reduction is then to modify the initial data parameters $\dot{r}_0$, $r_0$, and $\Omega_0$ so that the second term in Eq.~\eqref{eq:omegadot_analytic}
vanishes, and therefore $e = 0$ in the
Newtonian limit.

\begin{figure*}[t]
  \centering
  \includegraphics[width=.9\textwidth]{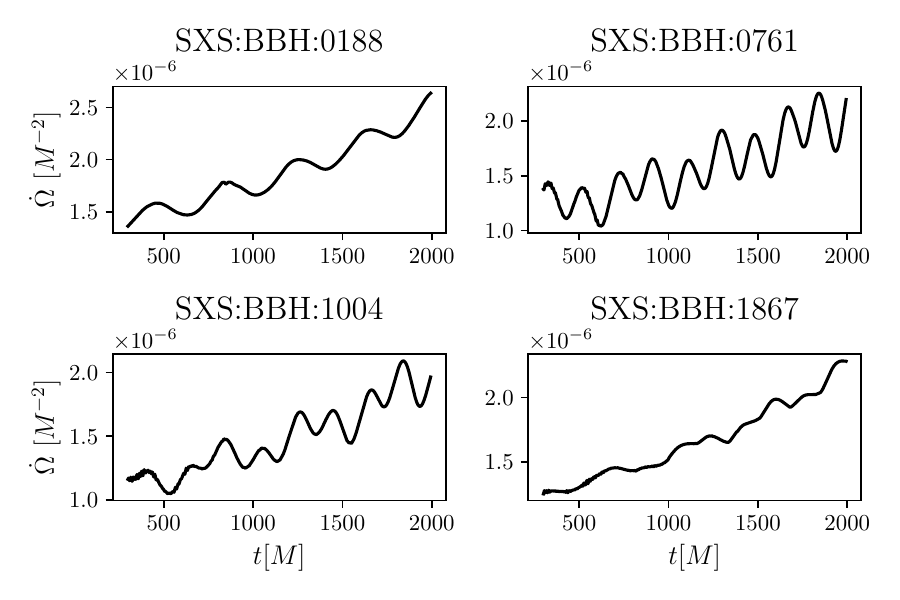}
  \begin{tabular}{ c c c c c }
    & $q$ & $e\times10^{4}$ & $\vec{\chi}_1$ & $\vec{\chi}_2$ \\
    \hline
    SXS:BBH:0188 & 7.187 & $1.609$ & (0, 0, 0) & (0, 0, 0) \\
    SXS:BBH:0761 & 2.0 & $1.923$ & (-0.63, -0.49, 0) & (0, 0, 0) \\
    SXS:BBH:1004 & 1.602 & $2.939$ & (-0.17, 0.41, -0.57) & (-0.49, -0.48, 0.21) \\
    SXS:BBH:1867 & 3.464 & $0.695$ & (0, -0.16, 0) & (-0.20, 0.17, -0.23) \\
\end{tabular}
  \caption{Some $\dot{\Omega}(t)$ trajectories from public SXS BBH simulations.
  The quantity $\dot{\Omega}(t)$ generally increases during inspiral, but both
  eccentricity and spin-spin precession can induce oscillations. Models such as
  that described in Sec.~\ref{sec:eccred} measure eccentricity by characterizing
  these oscillations. The spin-spin oscillations occur at about twice the
  frequency of the eccentricity-induced oscillations, so spin-spin effects could
  be accounted for either by fitting for them (i.e., the last term in
  Eq.~\eqref{eq:omegadot_impl}), or by removing them via a low-pass filter. In
  nonspinning simulations such as SXS:BBH:0188, only eccentricity-induced
  oscillations are present.}
 \label{fig:trajectories}
\end{figure*}

For curve fitting, we use a form of Eq.~\eqref{eq:omegadot_analytic} with
simplified prefactors for the sinusoidal terms,
\begin{align}
    \label{eq:omegadot_impl}
    \dot{\Omega}(t) = A\enclosepar{T_c - t}^{-11/8} + B\enclosepar{T_c -
                      t}^{-13/8}  \nonumber \\
  + \ C\cos\enclosepar{\omega t + a
  t^2 + \phi_0} - D\sin\enclosepar{ \bar{\alpha}(t) + \phi_s }, \\
    \label{eq:alphabardef}
    \cos\bar{\alpha}(t) \equiv \frac{\vec{S}_0 - \enclosepar{\vec{S}_0 \cdot
    \hat{L}_N}\hat{L}_N}{\enclosebar{\vec{S}_0 - \enclosepar{\vec{S}_0 \cdot
    \hat{L}_N}\hat{L}_N}} \cdot \hat{n}.
\end{align}
Here the first two terms describe radiation reaction \cite{Buonanno:2010},
and are equivalent to $S_\Omega(t)$ in
Eq.~\eqref{eq:omegadot_analytic}.
The final term in Eq.~\eqref{eq:omegadot_impl} is equivalent to the final term
in Eq.\eqref{eq:omegadot_analytic}, and
captures oscillations from
spin-spin interactions\footnote{It can be shown that
$\sin{2\bar{\alpha}} = 2 \sin(2\bar{\Omega} t + \gamma)$.}. For nonspinning
binaries, this term goes to zero and should be omitted from fitting. Our goal is
to fit the derivative of the orbital frequency, as obtained from an NR
simulation, to Eq.~\eqref{eq:omegadot_impl}. This fit has nine unknown
parameters to be determined: $A$, $B$, $C$, $D$, $T_c$, $\omega$, $a$, $\phi_0$,
and $\phi_s$.

From Eq.~\eqref{eq:omegadot_analytic} and the parameters found by fitting to Eq.~\eqref{eq:omegadot_impl}, eccentricity is estimated
as
\begin{align}
\label{eq:ecc_impl}
    e = \frac{C}{2\bar{\Omega} \omega},
\end{align}
following from the Newtonian definition of eccentricity and
Eq.~\eqref{eq:omegadot_analytic}. $\Omega_0$ is used as an approximation in place of $\bar{\Omega}$
for the following results.

The resulting updating formulae, as detailed in Ref.~\cite{Buonanno:2010}, are
\begin{subequations}
\label{eq:update}
\begin{align}
    \label{eq:update_adot}
    \Delta \dot{r}_0 = & \frac{C}{2\Omega_0}\cos\phi_0, \\
    \label{eq:update_omega0}
    \Delta \Omega_0 = & -\frac{C\omega}{4\Omega_0^2}\sin\phi_0.
    % \label{eq:update_r0}
    % \Delta r_0 = & -\frac{Cr_0\omega}{2\Omega_0 \enclosepar{\Omega_0^2 + \frac{2}{r_0^3}}}\sin\phi_0
\end{align}
\end{subequations}

In the above discussion, we choose to fix $r_0$ and update $\dot{r}_0$ and
$\Omega_0$. Alternatively, one can choose to fix $\Omega_0$ and update $r_0$ and
$\dot{r}_0$ instead, using a similarly derived update formula for $r_0$. One
variable is fixed in order to set the scale of the orbit, and the other two are
corrected to determine other features of the
orbit.

Here we collect certain symbol definitions for clarity. The variable $\Omega_0$
is the initial orbital frequency, an initial condition specified in the elliptic
solver for initial data, and updated through Eq.~\eqref{eq:update_omega0}. The
variable $\Omega(t)$ refers to the orbital frequency time series obtained by
numerical evolution, that is then fit to the form of
Eq.~\eqref{eq:omegadot_impl}. The variable $\omega$ is the frequency of
oscillations in $\Omega(t)$ that are induced by eccentricity, and is a fit
parameter in Eq.~\eqref{eq:omegadot_impl}. The variable $\bar{\Omega}$ is the
mean value of $\Omega$.

\subsection{Improved algorithm for eccentricity fitting}
\label{sec:new_improved_algorithm}

Overall, the method described in Sec.~\ref{sec:eccred}
works reasonably well, and has been used to reduce eccentricity for
all of the non-eccentric NR simulations in the SXS simulation
catalog~\cite{Boyle:2019}.  However, we have found that a key step in
the method, fitting NR trajectories to Eq.~\eqref{eq:omegadot_impl}, is
not very robust and is often the limiting factor that determines how
small an eccentricity can be obtained.  In particular,
standard nonlinear fitting methods sometimes fail to converge when
fitting to Eq.~\eqref{eq:omegadot_impl}, and even worse, the results of
the fit are sometimes extremely sensitive to small details such as
the time interval chosen for fitting.  These problems are related to
the large number of nonlinar fit parameters in Eq.~\eqref{eq:omegadot_impl},
and to the difficulties in choosing initial guesses for these parameters.

In this section, we propose some new techniques to mitigate these problems and
thus improve the robustness of eccentricity measurement. First, we detail a new
frequency-domain method for removing spin-spin oscillations from
$\dot\Omega(t)$. This method allows $\dot\Omega(t)$ to be fit to a simpler
function that has fewer fit parameters, and it provides a better initial guess
for the fit parameter $\omega$. We then briefly introduce the variable
projection algorithm\cite{Golub:1973} for solving nonlinear least squares
problems and show how this algorithm can reduce the number of
nonlinear parameters in the fit. Next, we discuss the choice of
initial guesses for the fit parameters, why the results of the previous method
are sensitive to this choice, and how this choice can be improved through the
use of variable projection and frequency-domain filtering.

\subsubsection{Frequency Domain Pre-processing}
\label{sec:fft}

\begin{figure*}[t]
    \centering
    \includegraphics[width=.9\textwidth]{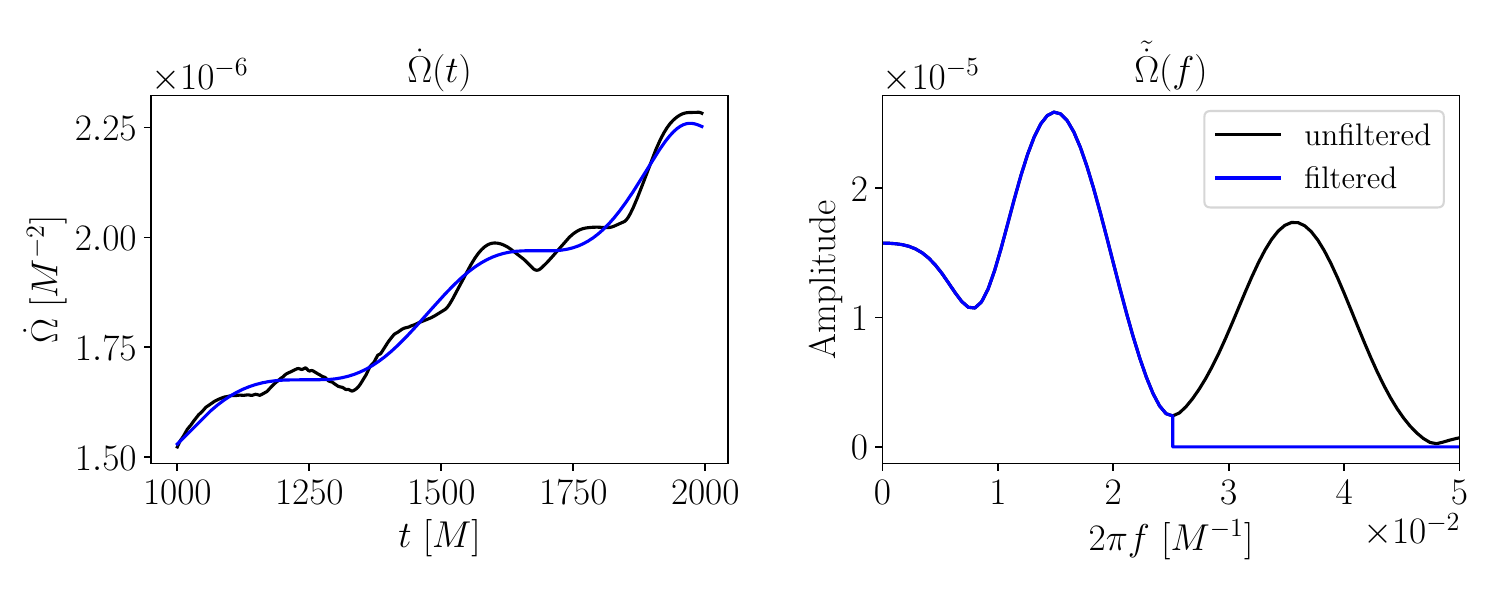}
    \caption{The $\dot{\Omega}$ time series from \texttt{SpEC} BBH run
    SXS:BBH:1867 and its corresponding frequency spectrum. The black curves
    correspond to the original data and the blue curves correspond to the data
    after removing high frequency content as outlined in Sec.~\ref{sec:fft}. The
    dominant frequency corresponds to eccentricity-induced oscillations, and can
    be used as an initial guess for the fit parameter $\omega$. Since the
    computed updates in Eq.~\eqref{eq:update} only depend on the dominant
    frequency, higher frequency content can be removed to simplify fitting (by
    dropping the last two terms in Eq.~\eqref{eq:omegadot_varpro}) without
    significantly impacting the relevant fit parameters. }
   \label{fig:spin_filter_FFT}
\end{figure*}

The last term in Eq.~\eqref{eq:omegadot_impl} describes oscillations caused by
spin-spin interactions. However, the parameters in that term do not enter into
the measurement of the eccentricity or the updating formulas for $\Omega_0$ and
$\dot{r}_0$ --- that term and its two fit parameters are present solely to model
an effect that we subtract out. We can take advantage of this to simplify
Eq.~\eqref{eq:omegadot_impl}. The last term in Eq.~\eqref{eq:omegadot_impl}
describes oscillations in $\dot{\Omega}(t)$ with roughly twice the frequency of
the oscillations caused by eccentricity (recall that $\omega$, $\Omega_0$, and
$\bar\Omega$ are all equal in the Newtonian limit for small eccentricity).
Figure~\ref{fig:trajectories} shows a small selection of $\dot{\Omega}(t)$
trajectories extracted from simulations in the SXS public catalog
\cite{Boyle:2019, SXSCatalog}. Notice that simulations SXS:BBH:0761,
SXS:BBH:1004, and SXS:BBH:1867, which have significant precessing spins, have a
much more complicated $\dot{\Omega}(t)$ with higher frequencies than does
SXS:BBH:0188, which has no spins. The higher frequency of spin-spin interactions
suggests that we can remove the spin-spin term in Eq.~\eqref{eq:omegadot_impl}
if we first apply an appropriate low-pass filter to the frequency spectrum of
$\dot{\Omega}(t)$.

To perform a low-pass filter, we first compute the power spectrum of
$\dot{\Omega}(t)$. To do this, the time series $\dot{\Omega}(t)$ must be
preprocessed before taking a Fourier transform. This is because
$\dot{\Omega}(t)$ is nonperiodic and its mean value increases over the course of
inspiral. Here, we detail the preprocessing steps we have used.

First, we must account for the average increase in $\dot{\Omega}$ over time. This
trend is caused by radiation reaction, and we can approximate it as a linear
process because of the short timescale of the part of the signal that we will
fit to. Thus we fit a linear trend to $\dot{\Omega}(t)$ and subtract it off. We
then apply a window function to the interval of the signal being fit to mitigate
the Gibbs phenomenon \cite{Wilbraham:1848}. Empirically we find that a Hamming
window \cite{Harris:1978} is particularly effective in preserving the underlying
signal structure. The resulting time series is then zero-padded on both sides to
increase frequency-domain resolution. For a time series with time steps of size
${}\sim 10^{-1} M$, we note that ${}\sim 10^4$
zeros on each end gives a frequency resolution ${}\sim 10^{-3} \ M^{-1}$, which
we find is sufficient to resolve important features in the frequency domain.
Figure~\ref{fig:spin_filter_FFT} shows an $\dot{\Omega}$ trajectory extracted
from an SXS public catalog simulation~\cite{Boyle:2019} and transformed into the
frequency domain using the process described above.

Now we turn to the low-pass filter, which we use to remove spin-spin
oscillations in $\dot{\Omega}(t)$ and simplify Eq.~\eqref{eq:omegadot_impl} by
dropping the final term. In Fig.~\ref{fig:spin_filter_FFT}, the spin-spin
oscillations correspond to the peak at $2\pi f \sim 3.5\times 10^{-2} \ M^{-1}$
in the right panel. To perform the low-pass filter, we choose a cutoff frequency
at the first local minimum that occurs after the dominant frequency peak. In
Fig.~\ref{fig:spin_filter_FFT}, this cutoff frequency occurs at $2\pi f \sim
2.5\times 10^{-2} \ M^{-1}$. We then set $\tilde{\dot{\Omega}}$ to zero for all
frequencies greater than this cutoff. The new spectrum is then inverse
transformed to give a filtered time domain $\dot{\Omega}(t)$ usable for fitting.
This filtered $\dot{\Omega}(t)$ is shown as the blue curve in the left panel of
Fig.~\ref{fig:spin_filter_FFT}. We find that fitting to the filtered
$\dot{\Omega}(t)$, ignoring the final term in Eq.~\eqref{eq:omegadot_impl},
produces approximately the same value of eccentricity as fitting to the original
$\dot{\Omega}(t)$ and keeping the final term in Eq.~\eqref{eq:omegadot_impl}.

In addition to removing the spin-spin oscillations, the low-pass filter
also removes higher overtones in $\dot{\Omega}$ and also high-frequency
numerical noise, both of which are unmodeled by Eq.~\eqref{eq:omegadot_impl}
and can interfere with the robustness of fitting.
Thus the low-pass filter technique improves the
eccentricity measurement algorithm on several fronts.

\subsubsection{Initial guesses for fit parameters}
\label{sec:initial-guesses-fit}

The nonlinear least squares methods used for fitting $\dot{\Omega}(t)$
to Eq.~\eqref{eq:omegadot_impl} are iterative: Initial guesses for the
nonlinear fit parameters are provided, and the method refines those
initial guesses multiple times until convergence is achieved.  Care
must be taken to choose accurate initial guesses, because inaccurate
initial guesses can lead to convergence in local minima. Figure~\ref{fig:local_min}
shows slices in the solution space of Eq.~\eqref{eq:omegadot_impl} for
the parameter $\omega$. The cost function is computed for an
analytically generated sample dataset, and each curve shown uses a
different length in time of sample data.  For parameters $\omega$ and
$a$ (not shown), both of which enter as cosine arguments, there are
many local minima near the global minimum. Moreover, the
larger the time interval over which $\dot{\Omega}(t)$ is fit, the more
local minima are present, and the closer they are to the desired
global minimum. This trend means that counterintuitively, fitting over
a longer inspiral does not necessarily translate to finding a more
accurate solution. However, inclusion of at least one full orbital
period is crucial for determining an optimal value of $\omega$.

\begin{figure}[t]
    \centering
    \includegraphics[width=\columnwidth]{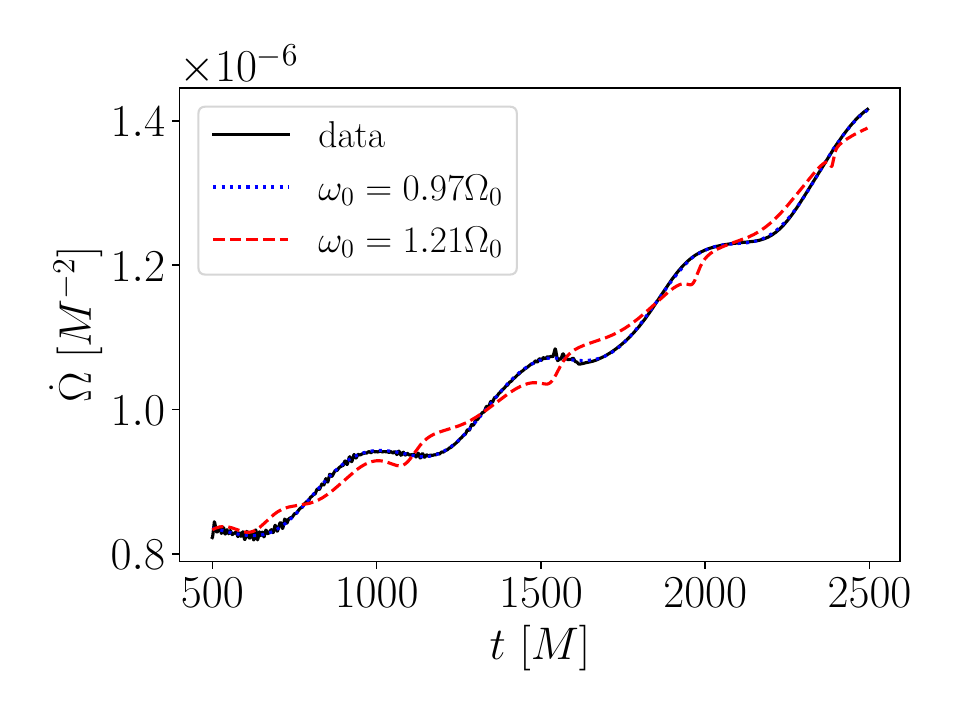}
    \caption{An $\dot{\Omega}$ time series for an equal mass, nonspinning BBH
    numerically simulated using \texttt{SpEC} (SXS:BBH:2085) and
    resulting curves of best fit found using variable projection for different
    initial guesses of $\omega$. In particular, the fit in blue (barely visible
    over the black curve) uses a guess for $\omega$ found from the dominant peak
    in the Fourier transform of $\dot{\Omega}$, as described in
    Sec.~\ref{sec:fft} and~\ref{sec:initial-guesses-fit}. A small change in the
    value of this guess can greatly impact the quality of the resulting fit, to
    the extent that primary features are not captured, as in the red fit curve.}
    \label{fig:bad_guess}
\end{figure}

Empirically, we observe that the guess for $\omega$ is the most
important factor in guaranteeing that  we converge to the global minimum.  The
typical total time interval used for fitting, roughly one to two
orbital periods, is too small to allow for high resolution of $T_c$ or
$a$, and neither $T_c$ nor $a$ appear in the formula for eccentricity
or the updating formulas. As Fig.~\ref{fig:bad_guess} shows, the value
of the $\omega$ guess has a large impact on the quality of the
resulting fits, regardless of the fitting technique used. A poor
guess for $\omega$ can result in a fit that entirely fails to capture
the primary oscillations in $\dot{\Omega}(t)$ caused by eccentricity.

For the previous method of eccentricity reduction described in
Sec.~\ref{sec:eccred}, the difficulty of choosing accurate initial guesses
typically prevents brute-force fitting to Eq.~\eqref{eq:omegadot_impl}. Instead,
the algorithm proceeds with a series of models and fits, where the first fit in
the series uses a model consisting only of a few terms in
Eq.~\eqref{eq:omegadot_impl} and therefore fewer parameters.  The next fit adds
another term and more parameters to the model, and uses the results of the
previous fit to provide initial guesses for the parameters.  This process
continues until the final fit uses all terms in Eq.~\eqref{eq:omegadot_impl}.
Each intermediate fit effectively functions as a search for an accurate initial
guess for one or more new parameters.

We employ three strategies to tackle the problem of initial guesses. The first
strategy, discussed above in Sec.~\ref{sec:fft}, is to simplify the fitting
function by filtering, so that the last term of Eq.~\eqref{eq:omegadot_impl},
and the corresponding parameters that need initial guesses, can be dropped.  The
second strategy, discussed in Sec.~\ref{sec:varpro} below, is to use a fitting
technique called \emph{variable projection}, which solves for a subset of the
parameters using \emph{linear} least squares and therefore does not require
initial guesses for them.  As we will show below, the use of variable projection
means that initial guesses are required for only $T_c$, $\omega$, and $a$. The
third strategy is used for the parameter $\omega$, which empirically we have
found is the one that needs the most accurate initial guess. We obtain the
initial guess for $\omega$ from the frequency spectrum $\tilde{\dot{\Omega}}$
that we computed in Sec.~\ref{sec:fft}. In particular, we choose $\omega$ as the
center of the initial peak in $\tilde{\dot{\Omega}}$. Care must be taken to
correctly identify this frequency. We use a standard peak-finding routine
(e.g.~\texttt{scipy.signal.find\_peaks} \cite{SciPy}) on the amplitude spectrum,
with a limit on the minimum amplitude allowed, and restrict the peak search to
local maxima in a bandwidth of $\Omega_0 \pm40\%$. For the simulation shown in
Fig.~\ref{fig:spin_filter_FFT}, $\Omega_0 = 1.69\times10^{-2}\, M^{-1}$, and the
dominant frequency in $\tilde{\dot{\Omega}}$ is $1.49\times10^{-2} \, M^{-1}$ in
the right panel of the figure. This frequency is in the search bandwidth and
would be chosen as the initial guess. Rarely, there may be no peak in this
range, or else multiple peaks of comparable amplitude, in which case no guess
can be extracted from $\tilde{\dot{\Omega}}$. In that case, we default to
choosing $\omega$ to be $0.8\Omega_0$, a value that was chosen by
trial-and-error. In the few cases we have seen of this failure mode, it
indicates that the eccentricity is too small to measure reliably. We find that
when an initial guess for $\omega$ can be obtained from the frequency spectrum,
it offers an improvement over the previous trial-and-error method, as shown in
Fig.~\ref{fig:bad_guess}.

Two other parameters, $T_c$ and $a$, require initial guesses.  The
guess for $T_c$ is approximated from the quadropole formula.  Because
$a$ is typically at a much smaller scale than the other fit
parameters, we find $a = 0$ to be an adequate initial guess.

\subsubsection{Variable Projection}
\label{sec:varpro}

As discussed in Sec.~\ref{sec:fit}, Eq.~\eqref{eq:omegadot_impl} is an
especially challenging nonlinear least squares problem. Much of this difficulty
can be removed by noting that Eq.~\eqref{eq:omegadot_impl} leads to
a separable least squares problem, that is, some of the parameters 
in the fit enter the model linearly while others are
nonlinear.

The ``best'' algorithm for separable problems
has been known for a long time,
since 1973~\cite{Golub:1973}, and is called variable projection.
It was implemented originally in a
Fortran code called \texttt{VARPRO}~\cite{golub1972}.
The idea is to
start with initial guesses only for the nonlinear parameters. Then
standard linear least squares solves for the linear parameters by the usual
analytic process. Next, an iterative nonlinear fitting routine updates
the nonlinear parameters with the linear parameters held fixed. The
whole procedure is then iterated until a suitable tolerance is achieved. The
clever part of the algorithm is that the Jacobian of the cost function
with respect to the nonlinear parameters that is used in the nonlinear
fitting has a dependence on the linear parameters. This is because the
nonlinear parameters depend implicitly on the linear
ones. Ref.~\cite{Golub:1973} worked out this contribution to the
Jacobian---it can be computed explicitly from the analytic solution of the
linear least-squares
problem using linear algebra techniques. In general, this
algorithm is never worse than brute-force nonlinear least squares
fitting, and often succeeds when brute force fails. A major reason for
this is the reduced dimensionality of the nonlinear part of the fitting.

In the case of Eq.~\eqref{eq:omegadot_impl},
the nonlinear parameters in the  model are
$\omega$, $T_c$, $a$, $\phi_0$, and $\phi_s$, while
$A$, $B$, $C$, and $D$ are the linear parameters.
To get the maximum benefit out of variable projection, one should
reparameterize the
model so that as many parameters as possible enter linearly. For example,
a term of the form $C\cos(\omega t + \phi_0)$ should be rewritten as
$C \cos\phi_0\cos(\omega t) - C \sin\phi\sin(\omega t)$. This avoids having
to treat $\phi_0$ as a nonlinear parameter.
We will do this  below in
recasting Eq.~\eqref{eq:omegadot_impl} to Eq.~\eqref{eq:omegadot_varpro}.

In this work, we have relied on a modern implementation of
\texttt{VARPRO}~\cite{oleary2013} in Matlab. We have translated this
code into Python so that it can use the nonlinear solvers available in
Scipy. This Python version is publicly available at Ref.~\cite{varpro}.

\subsubsection{New fitting formula for $\dot{\Omega}(t)$}
\label{sec:fit}

\begin{figure}[t]
    \centering \includegraphics[width=\columnwidth]{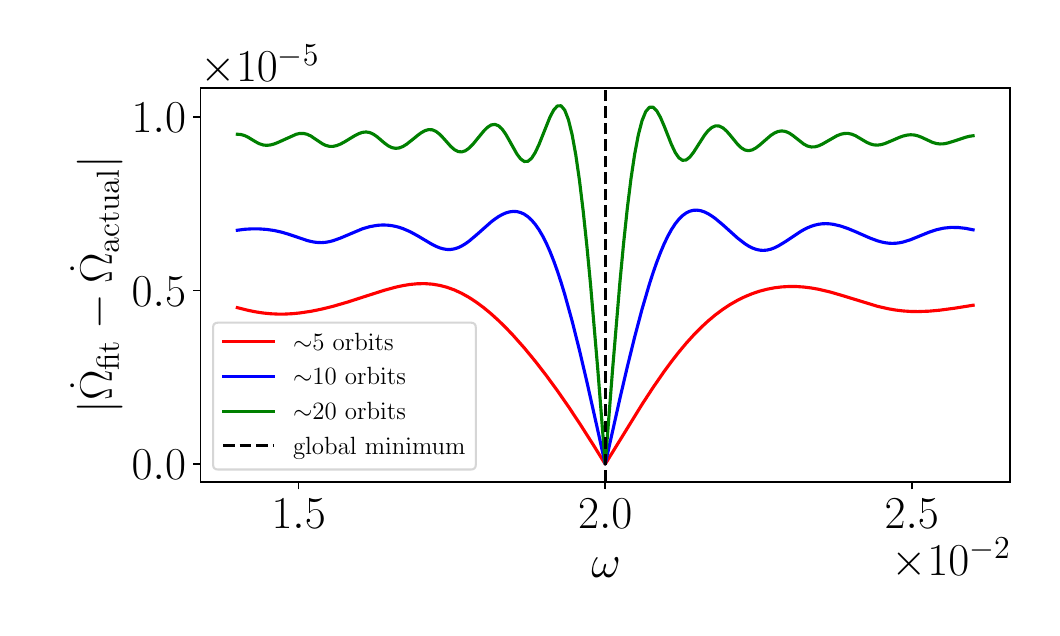}
    \caption{The cost function for Eq.~\eqref{eq:omegadot_impl} with a single
    parameter varied and all others left constant, calculated using trajectories
    of varying length. The residual curve of $\omega$, an argument for the
    sinusoidal components of Eq.~\eqref{eq:omegadot_impl}, contains several local
    minima near the global minimum. Furthermore, these local minima become
    closer and more numerous if time window length is increased. Optimizing
    Eq.~\eqref{eq:omegadot_impl} requires navigating around such local minima.}
    \label{fig:local_min}
\end{figure}

Eq.~\eqref{eq:omegadot_impl} can be rewritten as
\begin{align}
\label{eq:omegadot_varpro}
    \dot{\Omega}(t) &= \ A\enclosepar{T_c - t}^{-11/8} + B\enclosepar{T_c -
                        t}^{-13/8} \nonumber \\
                    & + \bar{C}_1\cos\enclosepar{\omega t + a t^2}  \nonumber \\
                    & - \bar{C}_2\sin\enclosepar{\omega t + a t^2} \nonumber \\
                    & -\bar{D}_1\cos\enclosepar{\bar{\alpha}(t)} -
                      \bar{D}_2\sin\enclosepar{\bar{\alpha}(t)},
\end{align}
where we have absorbed factors of $\cos\phi_0$, $\cos\phi_s$, etc., into new
parameters $\bar{C}_1$, $\bar{C}_2$, $\bar{D}_1$, and $\bar{D}_2$.  This
substitution eliminates two nonlinear parameters ($\phi_0$ and $\phi_s$) in
favor of two extra linear parameters, so that there are now six linear
parameters and three nonlinear parameters as opposed to four linear parameters
and five nonlinear parameters. Note that we have retained the
spin-spin terms, the ones with with coefficients $\bar{D}_1$ and $\bar{D}_2$,
for completeness, although in practice we can omit these terms as long
as we filter out the corresponding effects according to the procedure in Sec.~\ref{sec:fft}.

With variable projection, Eq.~\eqref{eq:omegadot_varpro} requires
guesses only for $T_c$, $\omega$, and $a$, so we have reduced the nonlinear fit
to three dimensions instead of the original nine. Note that only
the magnitude of $C$ (not $\bar{C}_1$ or $\bar{C}_2$) appears in the correction
formulae Eq.~\eqref{eq:update} and the eccentricity formula
Eq.~\eqref{eq:ecc_impl}, and $C$ is easily computed by $C^2 =
\bar{C}_1^2+\bar{C}_2^2$.

For all analyses in the following section, we drop the $B\enclosepar{T_c -
t}^{-13/8}$ term from Eqs.~\eqref{eq:omegadot_impl}
and~\eqref{eq:omegadot_varpro} for fitting. We find that the degeneracy in the
$A$ and $B$ parameters contributes some inconsistency to the fit solution, and
this term is difficult to resolve in a trajectory only 1-2 orbits long. It is
possible that fitting over a longer time span, where radiation reaction has a
bigger effect, might require keeping this term.

%==========================================================================
\section{Robustness Test}
\label{sec:results}

Here we compare the consistency of eccentricity measurements made both with and
without our improved fitting algorithm. We apply our methods first to an
analytic trajectory, then to simulations from the SXS public waveform
catalog~\cite{Boyle:2019, SXSCatalog}.

\begin{figure*}[t]
    \centering
    \includegraphics[width=.95\textwidth]{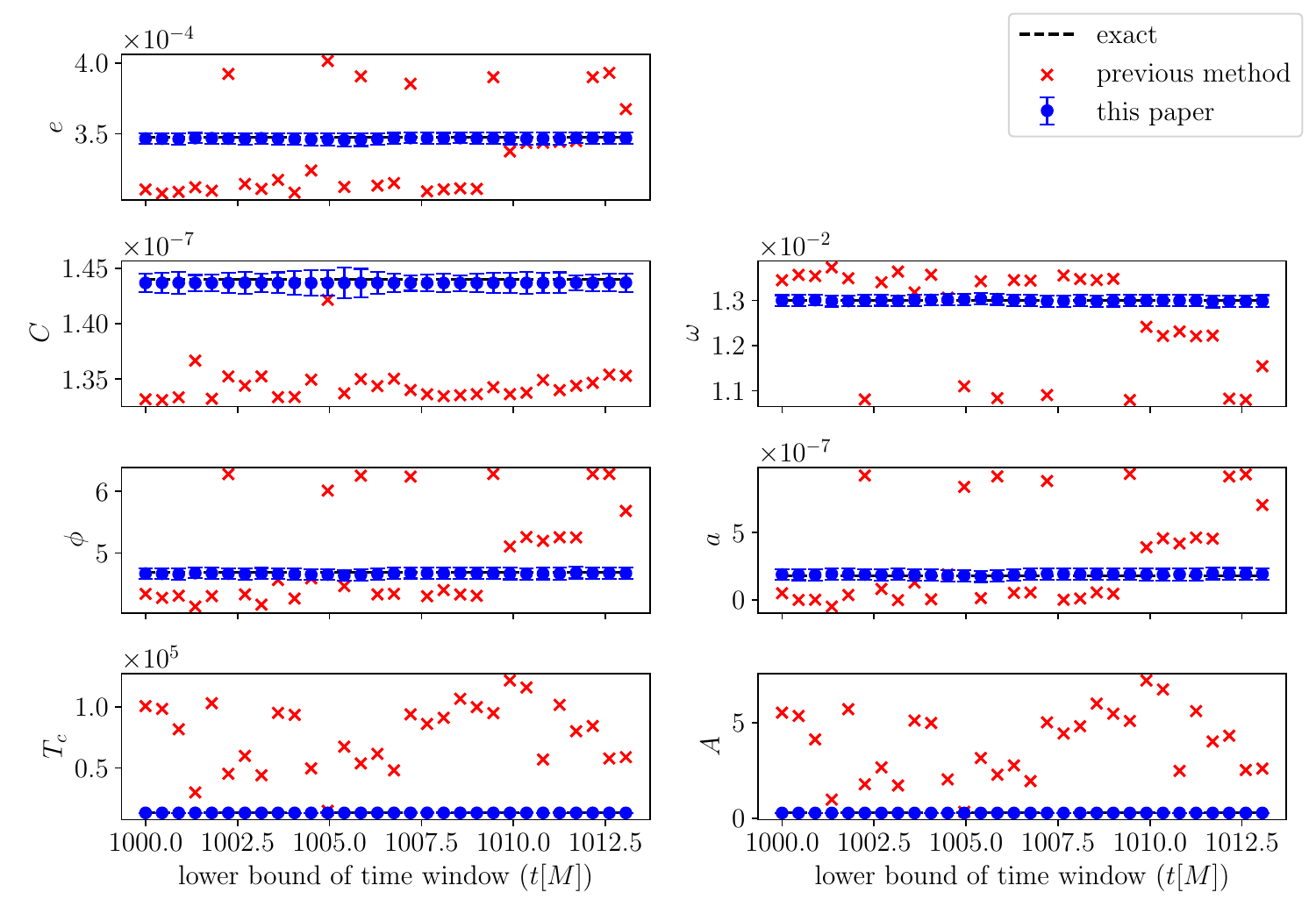}
    \caption{Best fit values for Eq.~\eqref{eq:omegadot_impl} (with the $B$ term
    dropped) and calculated eccentricity versus fitting window placement. We
    performed this method comparison using an analytic time series, not an NR
    simulation. For each set of window bounds, a nonlinear fit was performed on
    a dataset generated by fixing the parameters of Eq.~\eqref{eq:omegadot_impl}
    to exact numerical values and adding Gaussian noise to each point in the
    time series. The same window size, 900 $M$ (roughly two orbital periods),
    and bounds were used for each method. Fits labelled "this paper" were done
    with Eq.~\eqref{eq:omegadot_varpro} and initial guesses found using the
    method proposed in Sec.~\ref{sec:methods}, and found values of $\phi$ and
    $B$ are derived. Error bars for these points are computed as the
    square root of the corresponding diagonal element in the covariance matrix.
    Fits labelled "previous method" were done with the previous implementation
    for eccentricity reduction found in \texttt{SpEC}. Ideally, the result of
    the fit should have little to no dependence on the fit cutoff times, so the
    expected curve for each parameter is a horizontal line at the exact value.
    However, because the model has so many parameters that require guesses, the
    previous method is quite sensitive to the time window and often converges to
    an incorrect value.}
    \label{fig:analytic_window}
\end{figure*}

A key motivation for improving our algorithm is that the original method
described in Sec.~\ref{sec:eccred} is extremely sensitive to the time interval
$[t_\mathrm{min},t_\mathrm{max}]$ used for the fit. Consider
Fig.~\ref{fig:analytic_window}, which shows best fit parameters to an analytic
$\dot\Omega(t)$. The $\dot\Omega(t)$ used in Fig.~\ref{fig:analytic_window} is a
function that obeys Eq.~\eqref{eq:omegadot_impl} exactly, and is given as follows:
\begin{align}
\label{eq:analytic_traj}
    \dot{\Omega}(t) = \ & 0.287 \ (13000 - t)^{-11/8} + \\
    & (1.44\!\times\!10^{-7}) \cos(0.013t + (1.80\!\times\!10^{-7})t^2 + 4.68 ) \nonumber \\
    & + N(t) \nonumber
\end{align}
where $N(t)$ is noise from a Guassian distribution with a width of $10^{-8}$.
The red crosses in Fig.~\ref{fig:analytic_window} show the best fit parameters
of Eq.~\eqref{eq:omegadot_impl} for this $\dot\Omega(t)$, but as a function of
$t_\mathrm{min}$ (with $t_\mathrm{max}$ set such that $t_\mathrm{max} -
t_\mathrm{min}$ is the same for each point). For the red crosses, we
used the original fit method implemented in \texttt{SpEC}, the
\texttt{scipy.optimize.minimize}~\cite{SciPy} routine and a series of fits
incrementally adding terms from Eq.~\eqref{eq:omegadot_impl}. Small changes in
$t_\mathrm{min}$ (compared to the orbital period) can produce eccentricities
that vary by large amounts, sometimes even by a factor of 2 or larger.  This
sensitivity is effectively a source of noise that limits our ability to
accurately measure eccentricity, and this noise hampers the ability of the
eccentricity reduction procedure to converge to a small value of eccentricity.

\begin{figure*}[t]
    \centering
    \includegraphics[width=.95\textwidth]{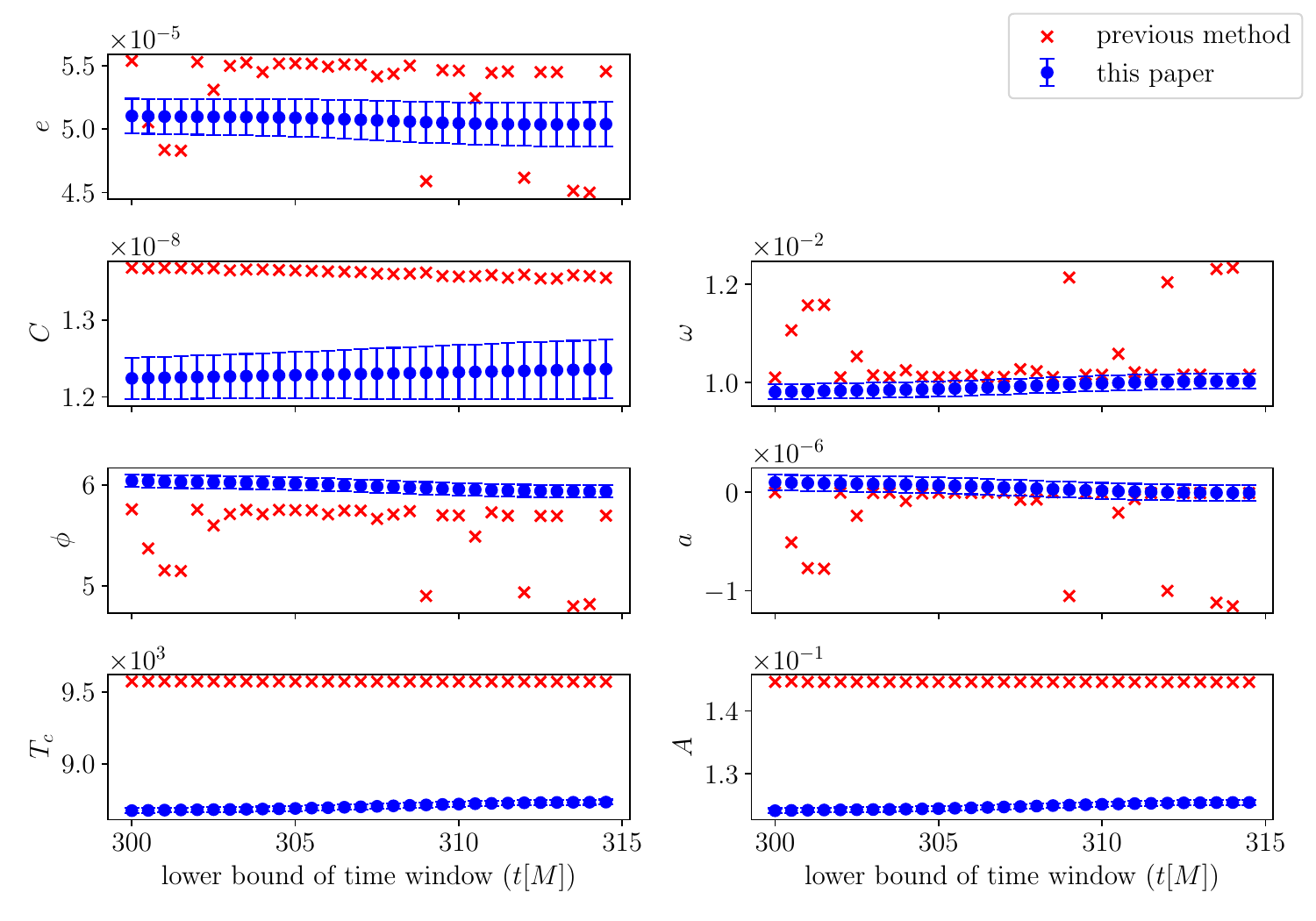}
    \caption{The same test as described in Fig.~\ref{fig:analytic_window}, using
    \texttt{SpEC} BBH trajectory SXS:BBH:0235 instead of an analytic dataset.
    Unlike in Fig.~\ref{fig:analytic_window}, an exact solution does not exist.
    For this test, $t_\mathrm{max} - t_\mathrm{min}$ was set to 1200 $M$,
    roughly two orbital periods.}
    \label{fig:window}
\end{figure*}

The blue circles in Fig.~\ref{fig:analytic_window} are the same as the red
crosses, except using the new techniques described in Sec.~\ref{sec:methods}.
When $t_\mathrm{min}$ is shifted, best fit values from the previous method often
jump discontinuously. Figure~\ref{fig:window} is the same as
Fig.~\ref{fig:analytic_window} except that $\dot\Omega(t)$ comes from unequal
mass, spinning \texttt{SpEC} BBH simulation SXS:BBH:0235. We see that even with
data from an NR simulation, the new method is significantly less sensitive to
$t_\mathrm{min}$ than the previous method. In addition,
Fig.~\ref{fig:analytic_window} shows that the new method tends to converge to
the correct solution more often. Overall, our proposed method gives considerable
improvement in both consistency and accuracy over the previously used algorithm.
Since variable projection is more successful than conventional algorithms at
converging to global solutions \cite{OLeary:2013}, the local minima highlighted
in Fig.~\ref{fig:local_min} are one likely source of the previously observed
inconsistency in measured eccentricity.

\begin{figure}[t]
    \centering
    \includegraphics[width=\columnwidth]{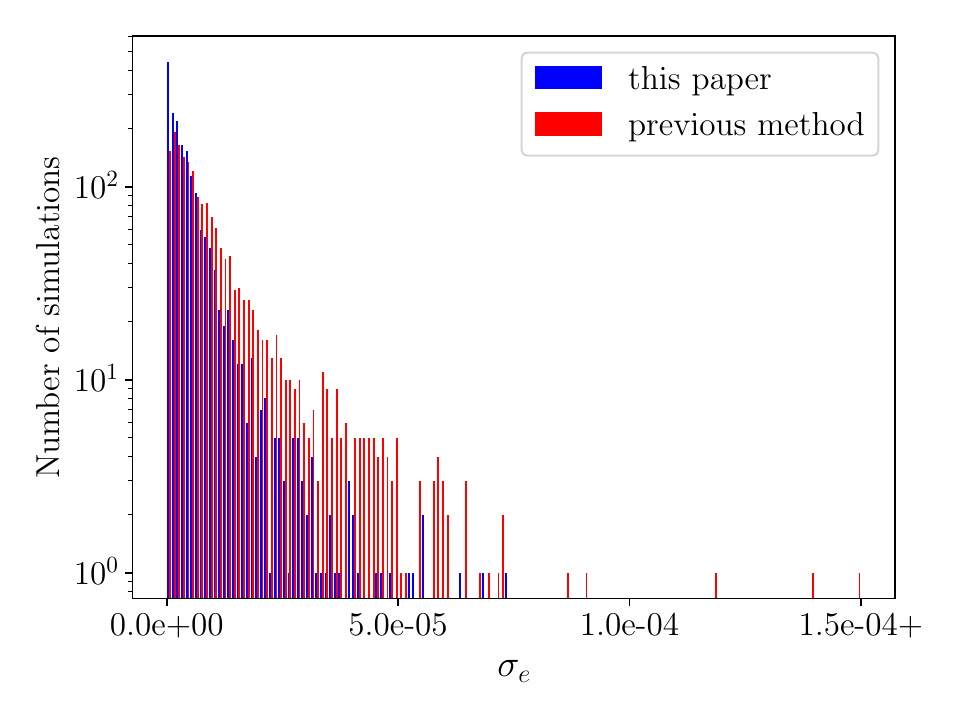}
    \caption{A histogram showing the standard deviation in fit $e$ with respect
    to fit time windows, $\sigma_e$, for simulations from the SXS public catalog
    with $e_{\mathrm{ref}} < 3\times10^{-3}$. For each simulation, the
    experiment in Fig.~\ref{fig:window} was repeated, from which $\sigma_e$ was
    then computed. We take this value as measure of fit consistency, and ideally
    it should be as close to 0 as possible. Bin width is $1\times10^{-6}$, and
    bin count corresponds to the number of simulations found to have $\sigma_e$
    within the bin range. In general, variation in measured $e$ when fitting
    with the techniques introduced in this paper is smaller than when using the
    current \texttt{SpEC} implementation. Note the logarithmic scale on the
    verical axis.}
    \label{fig:stability_histogram}
\end{figure}

One measure of the sensitivity of the eccentricity measurement to the time
interval $[t_{\mathrm min},t_{\mathrm max}]$ is to compute $\sigma_e$, the
standard deviation of $e$ as a function of $t_{\mathrm min}$ (i.e.~in the
appropriate subplot of Fig.~\ref{fig:window}). For Fig.~\ref{fig:window}, this
value is $\sigma_e=3.39\!\times\!10^{-6}$ for the previous method and
$\sigma_e=2.47\!\times\!10^{-7}$ for the method described in this paper. We now
repeat Fig.~\ref{fig:window} for all BBH simulations in the SXS public waveform
catalog. The catalog currently consists of about $2300$ precessing and
nonprecessing BBH simulations covering a broad parameter space spanning mass
ratios $1 \leq q \leq 10$ and spin magnitudes $0 \leq \chi_\mathrm{eff} \leq
0.998$ \cite{Boyle:2019}. We use simulations with a reference eccentricity
$e_{\mathrm{ref}} < 3\times10^{-3}$, totaling roughly $2200$ runs. For each
simulation in the catalog we compute $\sigma_e$, and we plot these values as a
histogram in Fig.~\ref{fig:stability_histogram}. We see a trend in that the best
fit $e$ with our proposed method is generally more consistent than with the
implementation that is currently used in \texttt{SpEC}.

Another issue with the previously used routine is that it occasionally fails to
converge to any solution during eccentricity reduction, often because of
reaching a maximum number of iterations without satisfying any exit criteria.
When this happens, initial data updates for eccentricity reduction cannot be
computed at all. Variable projection converges more often and with fewer fitting
iterations required. Several \texttt{SpEC} runs that previously failed because
the fit during eccentricity reduction failed to converge succeed when variable
projection is used in place of standard nonlinear least squares.

%==========================================================================
\section{Conclusion}
\label{sec:conclusion}

We have presented several improvements in an algorithm for reducing the orbital
eccentricity of binary simulations in NR. Unlike in Newtonian physics, it is not
straightforward in NR to specify initial orbital parameters that yield orbits
with zero eccentricity.  Instead, we use an iterative procedure, outlined in
Fig.~\ref{fig:flowchart}, in which eccentricity is estimated by doing
a short evolution.
If the estimated eccentricity is large, the initial data is corrected and the
evolution is restarted. This process is repeated until the measured eccentricity
is acceptably low.

The eccentricity estimation step of this iterative procedure involves fitting
the results of an NR simulation (in our case, the derivative of the orbital
frequency $\dot\Omega(t)$) to an estimator model, Eq.~\eqref{eq:omegadot_impl}.
Because the model has many parameters, some of which enter nonlinearly, the fit
requires a nonlinear least squares algorithm, which in turn requires accurate
initial guesses for the parameters so as to not fall into local minima of the
cost function. The model and its solution method are sensitive to
small details, and the previous method of eccentricity reduction used in
\texttt{SpEC}, described in Sec.~\ref{sec:eccred}, occasionally fails to
converge. It is sometimes possible to fix individual failures by hand, by
fine-tuning parameters of the algorithm such as initial guesses. However,
eccentricity reduction takes place as part as an automated pipeline that allows
a single user to run dozens or hundreds of BBH simulations at once
\cite{Boyle:2019}. When running hundreds of simulations, even infrequent
eccentricity reduction failures require significant human time to fix. A key
goal of the new method presented here is to eliminate or at least reduce these
failures.

We have found a number of new techniques that improve the reliability of best
fit solutions for the eccentricity estimator first derived in
Ref.~\cite{Buonanno:2010}. Our main improvements can be summarized as follows:
\begin{enumerate}
\item An initial guess for the $\omega_0$ fit parameter taken from the
$\dot{\Omega}$ frequency spectrum.
\item Use of variable projection for nonlinear least squares fitting and
reparameterization of Eq.~\eqref{eq:omegadot_impl} into
Eq.~\eqref{eq:omegadot_varpro} for fitting in order to fully take advantage of
variable projection. This effectively reduces the nonlinear least squares
model from nine to three dimensions.
\item Removal of high frequency content before fitting,
enabling the removal of an additional spin-spin precession term from the
model.
\item Removal of a higher-order radiation-reaction term from the model when appropriate, to reduce degeneracy in fit parameters.
\end{enumerate}

These methods are currently being integrated into the \texttt{SpEC}
code as part of the automated pipeline, and will also see future use in the
\texttt{SpECTRE} code \cite{spectrecode}. Note that although these techniques
have been implemented in \texttt{SpEC}, they are not specific to \texttt{SpEC}
and can be used in other NR codes. Further testing is needed to determine how
these improvements affect the number of iterations required
for  initial data correction
to achieve low eccentricity (e.g.~at or below $\mathcal{O}(10^{-4})$) and
what the lowest achievable eccentricity is.

A related problem to eccentricity reduction is measurement of larger
eccentricities and tuning of NR parameters to achieve a desired eccentric orbit.
Just as eccentricity is iteratively reduced as described in
Sec.~\ref{sec:eccred}, it can similarly be driven to a nonzero target value
using an eccentricity estimator requiring a nonlinear fit. The model presented
in Ref.~\cite{Buonanno:2010} is derived in the limit of small $e$, so
$\mathcal{O}(e^2)$ terms are dropped from Eq.~\eqref{eq:Newtonian_omega}. For
large eccentricity, a more general functional form is used, and an additional
parameter (the mean anomaly) must be specified. Application of the techniques
introduced here has not been fully explored in this case. We note that Ref.\
\cite{Knapp:2024} introduces a new technique for larger eccenctricities.

All testing presented here was done using BBH simulations. However,
the eccentricity reduction
method used here is not limited to black holes. The method as described in
Sec.~\ref{sec:eccred} is used in \texttt{SpEC} for BNS and BHNS simulations, so
the improvements found here apply to those simulations as well.

%==========================================================================
\begin{acknowledgments}

This work was supported in part by
the Sherman Fairchild Foundation, by NSF
Grants PHY-2207342 and OAC-2209655 at Cornell, and by
NSF Grants PHY-2309211, PHY-2309231, and OAC-2209656 at Caltech.
Computations were performed at Caltech using the Wheeler cluster
and Resnick HPC Center.

\end{acknowledgments}

%%%%%%%%%%%%%%%%%%%%%%%%%%%%%%%%%%%%%%%%%%%%%%%%%%%%%%%%%%%%%%%%%%%%%%%%%%%%%%%
% \section*{References}
%%%%%%%%%%%%%%%%%%%%%%%%%%%%%%%%%%%%%%%%%%%%%%%%%%%%%%%%%%%%%%%%%%%%%%%%%%%%%%%

\bibliography{References}

\end{document}